\documentclass[preprint]{aastex}

\slugcomment{Accepted by the \emph{Astrophysical Journal}}

\newcommand{\WMAP}{\textsl{WMAP}}

\newcommand{\COBE}{\textsl{COBE}}

\newcommand{\uK}{\ensuremath{\mu{\rm K}}}

\newcommand{\Gpc}{\textrm{Gpc}}

\newcommand{\ttponel}{\ensuremath{220.1\pm 0.8}}
\newcommand{\ttponeh}{\ensuremath{74.7 \pm 0.5}}
\newcommand{\ttponehsq}{\ensuremath{5583 \pm 73}}
\newcommand{\tttonel}{\ensuremath{411.7\pm 3.5}}
\newcommand{\tttoneh}{\ensuremath{41.0 \pm 0.5}}
\newcommand{\tttonehsq}{\ensuremath{1679 \pm 43}}
\newcommand{\ttptwol}{\ensuremath{546\pm 10}}
\newcommand{\ttptwoh}{\ensuremath{48.8 \pm 0.9}}
\newcommand{\ttptwohsq}{\ensuremath{2381 \pm 83}}

\newcommand{\tetonel}{\ensuremath{137\pm 9}}
\newcommand{\tetonehsq}{\ensuremath{-35\pm 9}}
\newcommand{\teptwol}{\ensuremath{329\pm 19}}
\newcommand{\teptwohsq}{\ensuremath{105\pm 18}}

\newcommand{\tthtwo}{\ensuremath{0.426\pm 0.015}}
\newcommand{\tththree}{\ensuremath{0.42\pm 0.08}}
\newcommand{\tehonehalf}{\ensuremath{0.33\pm 0.10}}

\newcommand{\dlphione}{\ensuremath{0.265\pm 0.006}}
\newcommand{\dlphithreehalfs}{\ensuremath{0.133\pm 0.007}}
\newcommand{\dlphitwo}{\ensuremath{0.219\pm 0.008}}
\newcommand{\dlphithree}{\ensuremath{0.299\pm 0.005}}
\newcommand{\dlpredictedlthreehalfs}{\ensuremath{409\pm 4}}
\newcommand{\dlpredictedltwo}{\ensuremath{533\pm 5}}
\newcommand{\dlpredictedlthree}{\ensuremath{809\pm 7}}
\newcommand{\acousticscale}{\ensuremath{300\pm 3}}
\newcommand{\acousticangle}{\ensuremath{0.601\pm 0.005\arcdeg}}
\newcommand{\dAdec}{\ensuremath{13.7\pm 0.4}}
\newcommand{\rsdec}{\ensuremath{143\pm 4}}

\newcommand{\mcmcttponel}{\ensuremath{219.8\pm 0.9}}
\newcommand{\mcmctttonel}{\ensuremath{410.0\pm 1.6}}
\newcommand{\mcmcttptwol}{\ensuremath{535\pm 2}}
\newcommand{\mcmctetonel}{\ensuremath{151.2\pm1.4}}
\newcommand{\mcmcteptwol}{\ensuremath{308.5\pm1.3}}
\newcommand{\mcmcttponehsq}{\ensuremath{ 5617\pm 72}}
\newcommand{\mcmctttonehsq}{\ensuremath{1647\pm 33}}
\newcommand{\mcmcttptwohsq}{\ensuremath{2523\pm 49}}
\newcommand{\mcmctetonehsq}{\ensuremath{-45\pm 2}}
\newcommand{\mcmcteptwohsq}{\ensuremath{117\pm 2}}

\newcommand{\zreion}{\ensuremath{20}}

\begin{document}
\title{First Year \textsl{Wilkinson Microwave Anisotropy Probe}
(\WMAP\altaffilmark{1})
Observations:\\ Interpretation of the TT and TE Angular Power Spectrum Peaks} 

\author{
L. Page\altaffilmark{2},
M. R. Nolta\altaffilmark{2},
C. Barnes\altaffilmark{2},
C. L. Bennett\altaffilmark{3},
M. Halpern\altaffilmark{4},
G. Hinshaw\altaffilmark{3},
N. Jarosik\altaffilmark{2}, 
A. Kogut\altaffilmark{3}, 
M. Limon\altaffilmark{3,9}, 
S. S. Meyer\altaffilmark{5},
H. V. Peiris\altaffilmark{6},
D. N. Spergel\altaffilmark{6},
G. S. Tucker\altaffilmark{7,9}, 
E. Wollack\altaffilmark{3},
E. L. Wright\altaffilmark{8}
}

\altaffiltext{1}{{\WMAP} is the result of a partnership between Princeton
University and the {\sl NASA} Goddard Space Flight Center. Scientific guidance
is provided by the {\WMAP} Science Team.}
\altaffiltext{2}{Dept. of Physics, Princeton University, Jadwin Hall, Princeton, NJ 08544}
\altaffiltext{3}{Code 685, Goddard Space Flight Center, Greenbelt, MD 20771}
\altaffiltext{4}{Dept. of Physics and Astronomy, University of British Columbia, Vancouver, BC  Canada V6T 1Z1}
\altaffiltext{5}{Depts. of Astrophysics and Physics, EFI and CfCP, University of Chicago, Chicago, IL 60637} 
\altaffiltext{6}{Dept. of Astrophysical Sciences, Princeton University, Princeton, NJ 08544}
\altaffiltext{7}{Dept. of Physics, Brown University, Providence, RI 02912}
\altaffiltext{8}{UCLA Astronomy, PO Box 951562, Los Angeles, CA 90095-1562}
\altaffiltext{9}{National Research Council (NRC) Fellow}
\email{page@princeton.edu}


\begin{abstract}

The CMB has distinct peaks
in both its temperature angular power spectrum (TT) and
temperature-polarization cross-power spectrum (TE).
From the {\WMAP} data we find the first
peak in the temperature spectrum at $\ell=\ttponel$
with an amplitude of $\ttponeh\,\uK$;
the first trough at $\ell=\tttonel$ 
with an amplitude of $\tttoneh\,\uK$; and
the second peak at $\ell=\ttptwol$
with an amplitude of $\ttptwoh\,\uK$.
The TE spectrum has 
an antipeak at $\ell=\tetonel$ 
with a cross-power of $\tetonehsq\,\uK^2$,
and a peak at $\ell=\teptwol$
with cross-power $\teptwohsq\,\uK^2$.
All uncertainties are $1\sigma$ and
include calibration and beam errors.

An intuition for how the data determine the cosmological
parameters may be gained by limiting one's attention to a subset of parameters
and their effects on the peak characteristics.
We interpret the peaks in the context of 
a flat adiabatic $\Lambda$CDM model with 
the goal of showing how the cosmic baryon
density, $\Omega_bh^2$, matter density, $\Omega_mh^2$, scalar index, $n_s$,
and age of the universe are encoded in their positions and amplitudes.
To this end, we introduce a new scaling relation for the TE antipeak-to-peak
amplitude ratio and recompute known related scaling relations for the TT 
spectrum in light of the {\WMAP} data.
From the scaling relations, we show that \WMAP's tight bound
on $\Omega_bh^2$ is intimately linked to its robust detection
of the first and second peaks of the TT spectrum.
  
\end{abstract}

\section{Introduction}

{\WMAP} has mapped the cosmic microwave background (CMB) 
temperature anisotropy over the 
full sky with unprecedented accuracy \citep{bennett/etal:2003}.
The temperature angular power spectrum
\citep[TT,][]{hinshaw/etal:2003}
and the temperature-polarization cross-power spectrum
\citep[TE,][]{kogut/etal:2003}
derived from those
maps have a number of characteristic features.
It is these features, and our ability
to predict them, that make the anisotropy such a powerful tool 
for cosmology. Computer programs like CMBFAST
\citep{seljak/zaldarriaga:1996} efficiently compute the
TT and TE power spectra for a wide variety of 
cosmological parameters. The model spectra are then compared to the data
to deduce the best fit parameters \citep{spergel/etal:2003}. The 
distinctiveness and accuracy of the measured spectrum determines 
the degree to which the parameters may be distinguished.

While the parameters of cosmological models are ultimately found by 
maximizing the likelihood of the data given a model, and the 
validity of a model is assessed by the goodness of fit
\citep{spergel/etal:2003}, this analysis by itself provides no model
independent assessment of the features of the angular power spectrum. 
Both intuition and calculational simplicity fall by the wayside. In this 
paper we focus on the peaks of the TT and TE spectra.
There are four reasons to consider just these particular characteristics.
The first is that through an examination of the peaks, one can gain 
an intuition for how the cosmological parameters are encoded in the 
TT and TE spectra. The second is that by determining how the peaks
depend on cosmological parameters, one may quickly assess how potential 
systematic errors in the angular power spectra affect some of the 
cosmological parameters. The third is that 
alternative models, or additions to the best fit model, may be 
easily compared to the data simply by checking to see if the 
alternative reproduces the peak positions and amplitudes. 
Finally, the peaks serve as a simple check for the sophisticated 
parameter fitting described in \cite{spergel/etal:2003} and 
\cite{verde/etal:2003}.
 
The paper is organized as follows. In \S\ref{sec:background} we give a 
brief overview of the CMB. In \S\ref{sec:peakvals} we 
give {\WMAP}'s best fit values for the positions in $\ell$-space
and amplitudes of the peaks. We then consider, in \S\ref{sec:interp},
the cosmological information that comes from the
positions and amplitudes of the TT and TE peaks. 
We interpret
the peaks in terms of the best fit model to 
the full {\WMAP} data set
\citep{spergel/etal:2003} and make direct
connections to the decoupling epoch. We conclude in \S\ref{sec:conclude}. 

\section{Overview\label{sec:background}}

The CMB TT spectrum may be divided into 
three regions depending on the characteristic
angular scale of its features.
In each region, a different physical process dominates.
The regions correspond to:
(a) Angular scales larger than the horizon size at decoupling
as observed today. These correspond to $\theta>2\arcdeg$ or, equivalently
$\ell<\ell_{dec}\approx 90$. The low $\ell$
portion is termed the Sachs-Wolfe plateau \citep{sachs/wolfe:1967}. 
In this region one observes the relatively unprocessed primordial 
fluctuation spectrum because patches of sky with larger separations 
could not have been in causal contact at decoupling. 
(b) The acoustic peak region, 
$0\fdg2<\theta<2\arcdeg$ or $90<\ell<900$, which is 
described by the physics
of a 3000~K plasma of number density
$n_e\approx 300~{\rm cm}^{-3}$ responding to 
fluctuations in the gravitational potential produced by the dark matter.
(c) The Silk damping tail \citep{silk:1968}, $\theta<0\fdg2$ or $\ell>900$, 
which is produced by
diffusion of the photons from the potential fluctuations and the washing
out of the net observed fluctuations by the relatively 
large number of hot and cold regions along the line of sight.  
The basic framework in which to interpret the temperature 
anisotropy has been known for over thirty years
\citep{peebles/yu:1970,sunyaev/zeldovich:1970}.  

It has long been recognized that the positions and amplitudes
of the peaks in the $90<\ell<900$ region could be used to
constrain the cosmological parameters in a particular model 
\citep{doroshkevich/zeldovich/sunyaev:1978, 
kamionkowski/spergel/sugiyama:1994,
jungman/etal:1996,jungman/etal:1996b}. Previous
studies \citep{scott/silk/white:1995, hancock/rocha:1997, 
knox/page:2000, weinberg:2000, cornish:2001, podariu/etal:2001,  
miller.c/etal:2002, odman/etal:2002, douspis/ferreira:2002, 
debernardis/etal:2002,
durrer/novosyadlyj/apunevych:2003} have steadily increased
our understanding of the peaks and their significance. 

A summary of pre-{\WMAP} determinations of the position and amplitude
of the first peak is given in Table 1. 

\section{Determination of Peak Characteristics\label{sec:peakvals}}

We determine the peaks and troughs with fits of
Gaussian and parabolic functions to the data. Through this process,
we compress a large data set to eight numbers. 
Such a compression is similar to specifying the  
cosmological parameters; though, it is considerably easier to compute and
directed more toward the intrinsic characteristics of the data as 
opposed to a cosmological model. 
 
For the TT spectrum, $\Delta T^2_\ell \equiv \ell(\ell+1)C_\ell/(2\pi)$,
we model the spectrum for $100<\ell<700$ as composed of a 
Gaussian peak, a parabolic trough, and a second parabolic peak
independent of any particular cosmological model.
Using a parabola to fit the first peak results in a higher $\chi^2$
and systematically underestimates the amplitude. Parabolas are adequate for the
other features.
The independent parameters are the amplitude and position of each peak
and trough, the width of the first peak, and a continuity parameter.
The latera recta\footnote{
The \textit{latus rectum} of a conic section is the length of the chord parallel
to the directix that passes through the focus.
For the parabolic curve $y = y_0 + (x-x_0)^2/w$ the value of the latus
rectum is $|w|$.
} of the trough and second peak are derived by constraining
the model spectrum to be continuous.
The junction between the first peak and the first trough
(first trough and second peak) is fixed to be where the first peak
(first trough) equals the continuity parameter.
In other words, the continuity parameter is the value of 
the angular power spectrum where the curves meet.
There are a total of eight parameters.
We denote the positions of the first peak, first trough, and second
peak as $\ell_1^{TT}$, $\ell_{1.5}^{TT}$, $\ell_2^{TT}$ respectively.
The amplitudes of the first and second peak are
$\Delta T^2_{1,TT}$ and $\Delta T^2_{2,TT}$. 

The peak parameters are found in three ways:
(a) with a direct non-linear fit based on the Levenberg-Marquardt algorithm
\citep{press/etal:NRIC:2e};
(b) with a Markov Chain Monte Carlo (MCMC) \citep{christensen/etal:2001}
of the peak parameters (not the cosmological parameters);
and (c) with peak-by-peak fits with a Gaussian 
fitting function. We have also separately fit the first peak with a
Gaussian with additional parameters for kurtosis and skewness.
All methods yield consistent results.
The values we quote are from the MCMC. 
All fitting is done with the unbinned angular spectra
using the full covariance matrix $\Sigma_{\ell \ell'}$ \citep{verde/etal:2003}.
The year one {\WMAP} calibration uncertainty is 0.5\% which 
is added in quadruture  
to the fitted peak amplitude uncertainty.
The fit to the TT peak models yields a minimum
$\chi^2/\nu=702/593=1.18$. 
While not as good a fit to the data 
as a CMBFAST-derived spectrum, the determinations of the peaks
are the same for all three methods and are consistent
with those of the best fit CMBFAST model. 
 
For the TE spectrum 
the data between $40<l<450$ are modeled as a piecewise-continous
composite of a parabolic antipeak and parabolic peak.
(We do not fit the reionization region, $\ell<20$).
Each parabola has a latus rectum, height, and position; however, the latus 
rectum of the antipeak is constrained by the requirement that
the antipeak and peak are continuously joined at the zero-crossing.
Thus there are just five free parameters.
We impose a prior on the latus rectum of the peak to be $<150$.
The TE Fisher matrix depends on $C^{TT}_\ell$, $C^{TE}_\ell$, and $C^{EE}_\ell$.
The TT spectrum is fixed to be a best-fit model,
and the small EE contribution is neglected.
The best fit model spectrum has $\chi^2/\nu=504/466=1.08$.

\begin{figure*}[tb]
\plotone{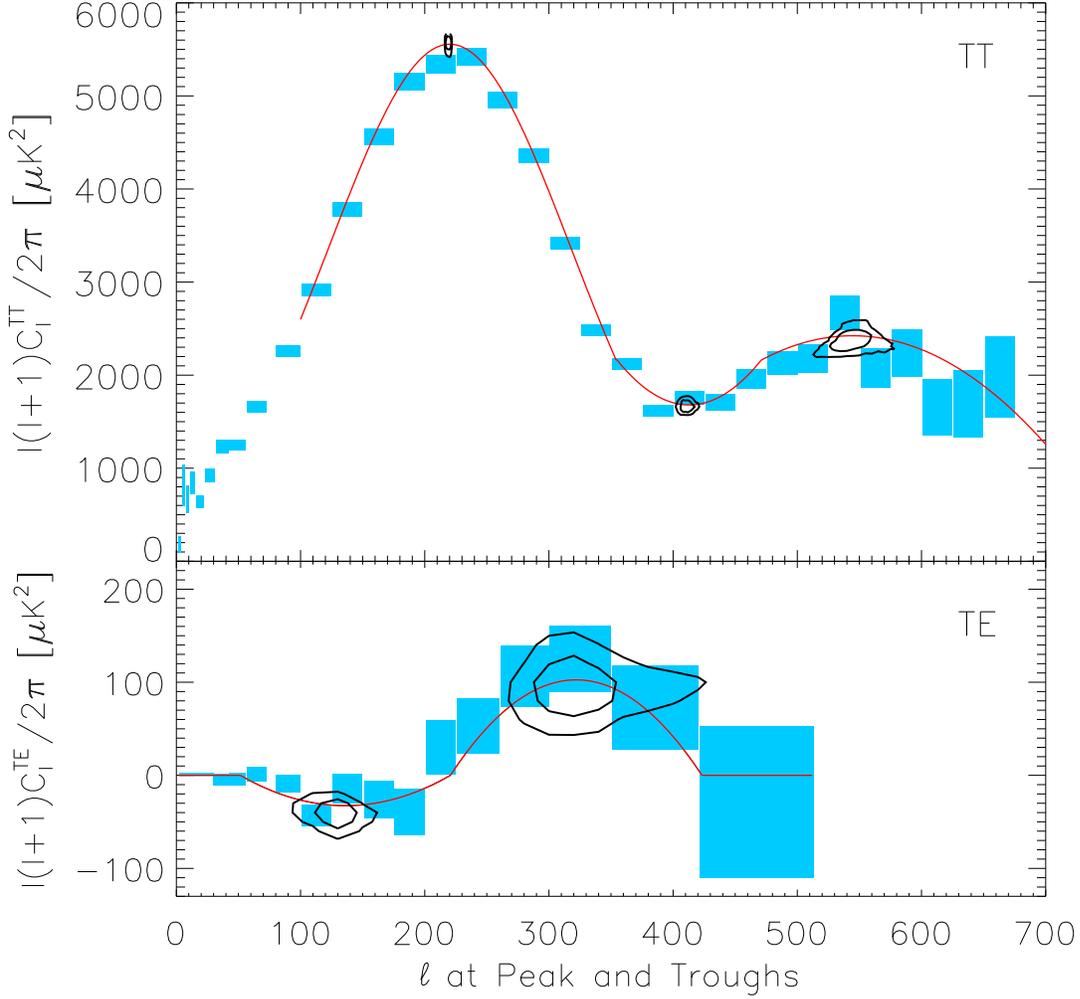}
\caption{The binned {\WMAP} data is shown in blue, 
the maximum likelihood peak model from the peak fitting functions in red,
and the uncertainty contours in black. 
The top panel shows the TT angular power spectrum. The
bottom panel shows the TE angular cross-power spectrum. For each peak 
or trough, the contours 
from the MCMC chains are multiplied by a uniform 
prior and so they are equal to contours of the a~posteriori 
likelihood of the data given the model. The
contours are drawn at 
$\Delta\chi^2 = 2.3$ and $6.18$ corresponding to 
$1\sigma$ and $2\sigma$.}
\label{fig:peakfit}
\end{figure*}

The parameters are summarized in Table~\ref{tab:peaks} along with 
the peak positions determined directly from the MCMC method.
These two completely distinct methods yield consistent results.
Figure~\ref{fig:peakfit} 
shows the binned TT and TE angular spectrum with the best fit peaks model
and the $1\sigma$ and $2\sigma$ contours for the peak and trough positions.
The amplitude of the TT trough and second
peak are separated by $>5\sigma$, leaving no doubt of the existence
of the second peak.

Also shown in Table~\ref{tab:peaks} are the peak determinations from
the full analysis using just the {\WMAP} data \citep{spergel/etal:2003}.
The agreement between the two is generally good, most of the values are
within $1\sigma$, all are within  $1.6\sigma$. This level of discrepancy
is expected because the peak fitting and the full analysis weight the data in 
very different ways. 

\section{Interpretation of Peaks and Troughs\label{sec:interp}}

We now interpret the peak characteristics in terms of a flat adiabatic
$\Lambda$CDM cosmological model. We focus on the
baryon density, $\omega_b=\Omega_bh^2$, 
the matter density, $\omega_m=\Omega_mh^2$
($\omega_m=\omega_b+\omega_c$, where $\omega_c$ is
the cold dark matter component),
and the
the slope of the primordial power spectrum, $n_s$.
The comparison of the peaks and troughs to the model is made at two 
levels. At the more general level, we use CMBFAST to derive 
simple scaling relations and show how the cosmological parameters are 
encoded in the peak characteristics. The aim is to build an 
intuition for the connection between the raw data and the deduced
parameters. This approach has been championed by \cite{hu/etal:2001};
this paper forms the basis for our presentation. The scaling 
relations hold for a wide range of parameters and may be derived 
without considering {\WMAP} data. However, we optimize them for the best 
fit {\WMAP} parameters (Table~\ref{tab:numvals}), so they are 
as accurate as possible. At a 
more detailed level, we show that by using the peak
values {\it and} uncertainties with the scaling relations, 
one obtains constraints and uncertainties on $\omega_b$, $\omega_m$, 
$n_s$, and the age of the universe that are consistent with the 
full analysis \citep{spergel/etal:2003}. 

\cite{seljak:1994} and \cite{hu/sugiyama:1995} showed that the physics
of the acoustic peaks may be understood in terms $\omega_b$, $\omega_m$, 
$n_s$,  $\tau$ (the optical depth to when 
the universe was reionized), and
$\theta_A$ (the angular scale of the sound horizon at decoupling).
Our approach follows that of \cite{hu/etal:2001} in that
we consider just the positions of the peaks and the ratios
of the peak amplitudes. For $\ell>40$,
the peak ratios are insensitive to the intrinsic amplitude
of the CMB spectrum and to $\tau$. The free electrons from reionization 
at $z\sim\zreion$ \citep{kogut/etal:2003} scatter CMB photons, 
thereby reducing the CMB fluctuations by nearly a constant factor
for multipoles $\ell > 40$. (The reionization also increases the TT spectrum
at $\ell<20$ due to Doppler shifts of the scatterers.) 
We therefore consider just $\omega_b$, $\omega_m$, $n_s$, and 
$\theta_A$, in the context of a flat $\Lambda$CDM model.

\subsection{The Position of the First Peak}

The acoustic peaks arise from adiabatic compression of the photon-baryon
fluid as it falls into preexisting wells in the gravitational potential.
These potential wells are initially the result of fluctuations
in some primordial field (e.g., the inflaton field in 
inflationary cosmology). The wells are enhanced by the dark 
matter ($\omega_c$), which is able to cluster following matter-radiation 
equality ($z_{eq}$, Table~\ref{tab:numvals}) because the dark matter,
for our purposes, does not scatter off the photons or baryons.
  
The first peak corresponds to the scale of the mode that has 
compressed once in the age of the universe at the time that 
the photons decoupled from the electrons at $z_{dec}$,
some \ensuremath{379^{+ 8}_{- 7} \mbox{ kyr}}
after the Big Bang \citep{spergel/etal:2003}.
The characteristic angular scale of the peaks is set by
\begin{equation}
\theta_A \equiv {{r_s(z_{dec})}\over{d_A(z_{dec})}},
\label{eq:theta_s}
\end{equation}
where $r_s$ is the comoving size of the sound horizon at 
decoupling and $d_A$ is the comoving angular size distance
to the decoupling surface.

The size of the sound horizon may be computed from the properties of
a photon-baryon fluid in an expanding universe. 
At decoupling it is given by \citep[B6]{hu/sugiyama:1995}:
\begin{equation}
r_s(z_{dec}) =
3997 \sqrt{{{\omega_\gamma}\over{\omega_m \omega_b}} }
\ln { \sqrt{1+R_{dec}}+\sqrt{R_{dec} + R_{eq}}\over 1 +\sqrt{R_{eq}} }
\label{eq:rss}
\end{equation}
in Mpc, where $R(z)=3\rho_b(z)/4\rho_\gamma(z)$ 
with $\rho_b$ the baryon density and $\rho_\gamma$ the photon density.
The redshift at matter-radiation equality is
\begin{equation}
1+z_{eq}={ 5464\,(\omega_m/0.135) \over 
(T_{\rm CMB}/2.725)^4(1+\rho_\nu/\rho_\gamma)}.
\end{equation} 
The ratio of neutrinos to photons is $\rho_\nu/\rho_\gamma=0.6851$
\citep{hannestad/madsen:1995}. Note that $r_s(z_{dec})$ depends only on the 
physical densities, $\omega_m$ 
and $\omega_b$, and not on $h$. (We assume $T_{\rm CMB}=2.725\,{\rm K}$ 
\citep{mather/etal:1999} 
and the number of neutrino species, $N_\nu=3$.)  Also, $r_s$ is independent 
of the curvature and cosmological constant densities, 
$\Omega_k$ and $\Omega_\Lambda$, because these are both late acting:
at high $z$ the universe evolves as though it is 
geometrically flat. 

The comoving angular size distance\footnote{We follow the naming
convention of \citet[13.29]{peebles:POPC} for $d_A$.
Common alternative names include ``proper motion distance'' 
\citep[e.g.,][14.4.21]{weinberg:GAC},
``angular diameter distance'' \citep[e.g.,][]{efstathiou/bond:1999},
and ``transverse comoving distance'' \citep[e.g.,][]{hogg:1999}.}
to the decoupling surface, $d_A$, for a flat geometry is
\begin{equation}
d_A = \int_0^{z_{dec}} 
{H_0^{-1}\,dz\over\sqrt{\Omega_r(1+z)^4 + \Omega_m(1+z)^3 + \Omega_\Lambda}},
\label{eq:add}
\end{equation}
where $\Omega_r$ is the current radiation density and 
$\Omega_\Lambda = 1 - \Omega_m - \Omega_r$ \citep{peebles:POPC}.
Neither $r_s$ nor $d_A$ depend on $n_s$.
The picture one has is that of measuring the angular size $\theta_A$ of a 
physical meter stick of size $r_s$ near the edge of the observable universe
with geodesics of the intervening geometry as encoded in $d_A$.

The position in $\ell$-space of the first peak is intimately linked to
$\theta_A$. In fact, in the full analysis \citep{spergel/etal:2003},
$\theta_A$ is one of the best determined fit parameters because of its
association with the first peak. To make the connection, one defines
a characteristic acoustic index, $\ell_A\equiv \pi/\theta_A$.
However, $\ell_A$ is not $\ell_1^{TT}$ because the 
potential wells that drive the compression, leading to the large variance
at $\ell_1^{TT}$, are dynamic: they respond to the matter and radiation
that falls into them. Additionally, there is no precise physical relation
between an angular scale and an associated $\ell$. 
One calibrates the relation with cosmological models, such as those 
from CMBFAST, to find a phase factor
$\phi_1$ that relates the two \citep{hu/etal:2001}. The general relation
for all peaks and troughs is
\begin{equation}
\ell_m^{TT}=  \ell_A(m-\phi_m),
\label{eq:peak_phase}
\end{equation}
where $m$ labels the peak number ($m=1$ for the first peak, $m=1.5$ for the
first trough, etc.). For a particular curvature density, $\Omega_k$, 
the $\phi_m$ depend weakly on $\omega_b$, and $\omega_m$.
For example, near the {\WMAP} values, changing $n_s$, $\omega_b$,
and $\omega_m$ by 10\% changes $\phi_1$ by 4.6\%, 0.5\%, and 1.1\%, 
respectively.
Because the $\phi_m$ differ by only $\approx25$\% between peaks, $\ell_A$
may be thought of as the characteristic scale of the peaks in $\ell$-space.
Effectively, the full analysis solves for the phase factors simultaneously
with the other parameters.

The considerations above are quite general and involve a good fraction
of the quantities that are of interest to cosmologists. We now show how the
{\WMAP} peak positions are related to, and in some cases determine, these
quantities. 

The position of the first peak is an essential ingredient of the mixture 
of properties that make the CMB such a powerful probe of the geometry 
of the universe \citep{kamionkowski/spergel/sugiyama:1994,jungman/etal:1996}.
However, the position alone does not determine the geometry.
With the full TT spectrum, the quantities that {\WMAP} measures 
particularly well, independent of $\Omega_k$, are $\omega_m$ and $\omega_b$ 
because both these quantities affect the spectrum at early times, 
before geometric effects shift the spectrum. 
Even if $\omega_m$ and $\omega_b$ are 
known, $\Omega_k$ and 
$\Omega_m$ (through $h$) may be traded off one another to give 
the same $\theta_A$. This
is called the ``geometric degeneracy'' \citep{bond/efstathiou/tegmark:1997,
zaldarriaga/spergel/seljak:1997}. The degeneracy is broken with either
prior knowledge of $h$ or $\Omega_m$ 
\citep[e.g.,][]{freedman/etal:2001,bahcall/etal:1999}.
For {\WMAP} if we impose a prior constraint of $h>0.5$ then
$\Omega_k=0.03\pm0.03$; without the prior,
\ensuremath{\Omega_k = -0.050^{+ 0.036}_{- 0.039}} \citep{spergel/etal:2003}.
In the following, we take the geometry to be flat and therefore consider
the peak positions as a function of $\omega_b$, $n_s$, and $\omega_m$ only.

\begin{figure*}[tb]
\plotone{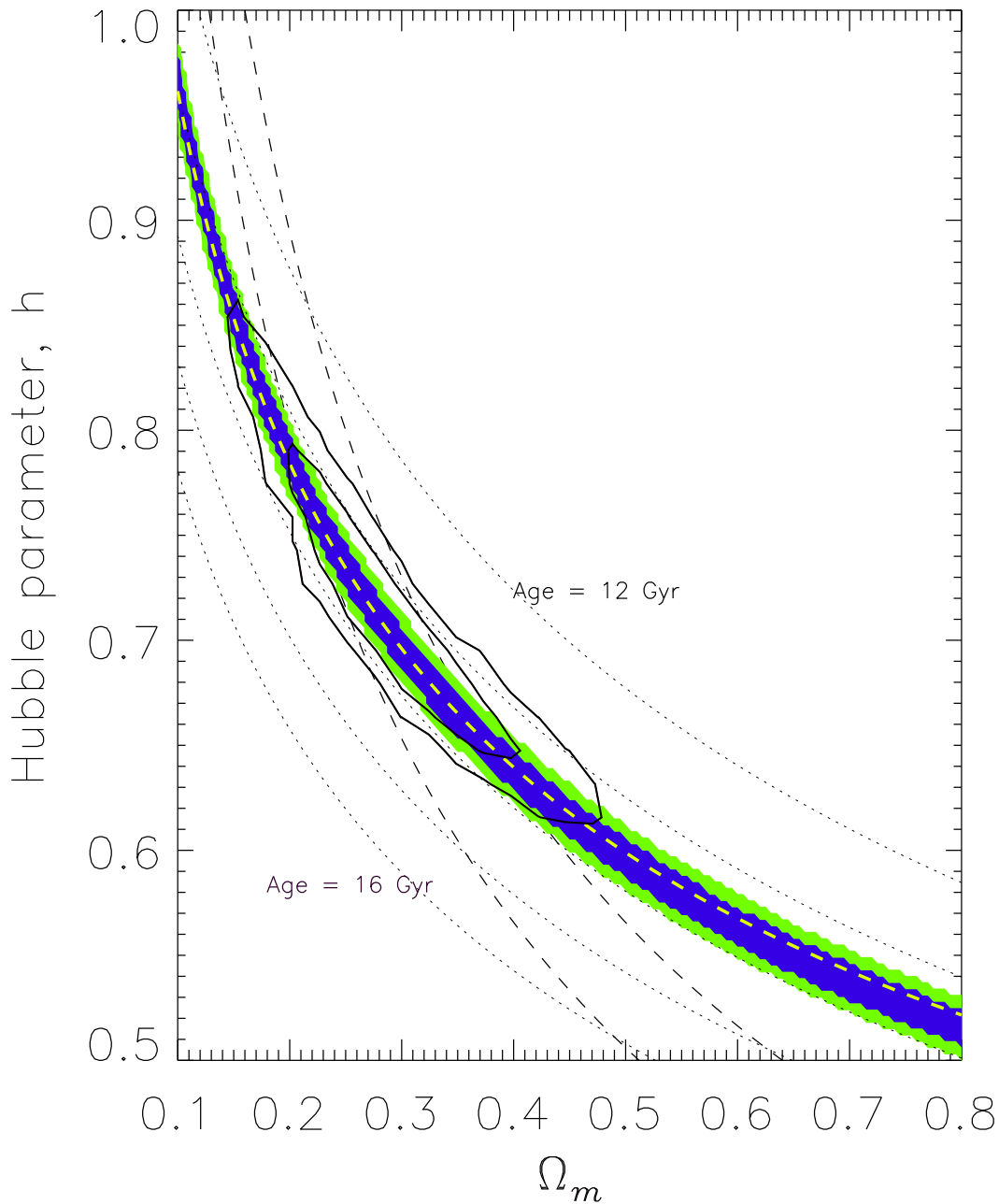}
\caption{\small The {\WMAP} data in the $\Omega_m-h$ plane. The thick
solid contours in black are at $\Delta\chi^2=-2.3, -6.18$ ($1\sigma$, $2\sigma$)
of the marginalized likelihood from the full analysis 
\citep{spergel/etal:2003}.
The filled region is the constraint from
the position of the first peak, with $\omega_b=0.023$ fixed.
In effect, it shows how $\Omega_m$ and $h$
must be related to match the observed position of the 
first peak in a flat geometry, or equivalently to match
the measured values of $\theta_A$.
Blue shows the $1\sigma$ region and green shows the $2\sigma$ region.
The dotted lines are isochrons separated by 1 Gyr. It is clear that
the {\WMAP} data pick out 13.6~Gyr for the age of the universe in the
flat, $w=-1$ case.
The dashed lines show the $1\sigma$ limits on $\omega_m$.
The dashed yellow line shows $\Omega_m h^{3.4} = {\rm const}$.
}
\label{fig:h_vs_omm}
\end{figure*}

With known $\omega_b$ and $\omega_m$,
the acoustic horizon size $r_s$ is particularly well determined.
From equation 2 we find $r_s=\rsdec~$Mpc.
When $n_s$ is included, we may in addition
determine the phases. From the fits in 
\cite{doran/lilley:2002} we obtain $\phi_1=\dlphione$.
The uncertainty is derived from the uncertainties of $n_s$, $\omega_b$,
$\omega_m$, and the quoted accuracy of the fitting function.
This combined with the measured position of the first peak at
$\ell_1^{TT}=\ttponel$ gives us an acoustic horizon scale of
$\ell_A=\acousticscale$. And, from $\ell_A$
we find that $\theta_A=\acousticangle$. In other words, we know 
the angular and physical sizes of structures 
on the decouping surface very well.
From equation~\ref{eq:theta_s}, we solve for the angular size distance
and find that $d_A = \dAdec\,\Gpc$. If we had naively tried to
compute $d_A$ from $h$ and $\Omega_m$ and their uncertainties
directly from equation~\ref{eq:add}, the resulting uncertainty
would have been larger. This, though, is not the correct 
procedure because the $h$ and $\Omega_m$ deduced from the 
CMB are correlated as shown in Figure~\ref{fig:h_vs_omm}.

The interplay between measuring the peak position, and thus $d_A$,
and measuring $\omega_m$ is at the root of \WMAP's ability to
determine $\Omega_m$ and $h$ separately 
\citep{bond/etal:1994, efstathiou/bond:1999}.
In particular, \cite{percival/etal:2002} show that $\theta_A$, which depends 
primarily on the first peak position, is the same along a line of 
constant $\Omega_mh^{3.4}$. In other words, this is how $\Omega_m$
and $h$ scale to keep $\theta_A$ constant. In Figure~\ref{fig:h_vs_omm}
we show the constraint from the measured first peak position.
For each coordinate pair in the plane, we compute $\ell_1^{TT}$
from equations~\ref{eq:theta_s}, \ref{eq:rss}, \ref{eq:add}, 
and \ref{eq:peak_phase}
and then assign the pair a likelihood derived from the measured distribution 
of $\ell_1^{TT}=\ttponel$. (We set $\omega_b=0.023$ for the calculation 
as discussed below.) The figure also shows the constraints from
$\omega_m$ and the likelihood contours from the full
{\WMAP} likelihood analysis. It is evident that the departure
from a fixed $\Omega_mh^2$ dependence in $\theta_A$ enables 
separating $\Omega_m$ and $h$. The separation is by no means complete
and constitutes one of the largest parameter degeneracy
for {\WMAP}; it precludes determining
$\Omega_\Lambda=1-\Omega_m$ to high precision with CMB data alone.
The independent determination of  $\Omega_m$ and $h$ will improve 
with time as the uncertainty on $\omega_m$ decreases.
Fortunately, galaxy surveys are sensitive to $\Omega_mh$ which breaks
the degeneracy when the data sets are combined \citep{spergel/etal:2003}.

It is fortuitous that in a flat geometry the position of the first
peak is directly correlated with the age of the universe
\citep{knox/christensen/skordis:2001,hu/etal:2001}. This
is seen most easily in the $\Omega_m-h$ plane in Figure~\ref{fig:h_vs_omm}.
With $w=-1$, the age depends only on 
the expansion time scale $h^{-1}$ and $\Omega_m$ and is given by
\begin{equation}
t_0 = 6.52h^{-1} (1 - \Omega_m)^{-1/2}
\ln\left({{1 + \sqrt{1-\Omega_m}}\over{\sqrt{\Omega_m}}}\right)
\ {\rm [Gyr]}
\label{eqn:t0}
\end{equation}
\citep{carroll/press/turner:1992}.
We show the isochrons for 12~Gyr through 16~Gyr. 
One observes that they overlap considerably
with the position constraint and the degeneracy lines from
equation~\ref{eq:add}.
For reasonable values of $\Omega_m$, the
age is seen to be in the neighborhood of 13.6 Gyr. The full
analysis \citep{spergel/etal:2003} gives \ensuremath{t_0 = 13.6 \pm 0.2 \mbox{ Gyr}} for
$w=-1$. Changing $\omega_b$ changes the position constraint
through changes in $r_s$ and $\phi_1$. If $\omega_b$ is increased 
from 0.024 to 0.025 ($1\sigma$), the position constraint line
shifts by $<\sigma/2$,
indicating that the plot is not particularly sensitive to changes in $\omega_b$.

To summarize this section, we have shown how the position of the first peak
is related to the cosmological parameters. Traditionally, one has used 
equation~\ref{eq:peak_phase}, the weak dependence of $\phi$ on
$\omega_b$ and $\omega_m$, $n_s\approx1$ and the measured position 
of the first peak
to deduce flatness. Instead, we assumed
flatness and use \WMAP's values of $n_s$, $\omega_b$ and $\omega_m$, 
to deduce a precise relation for the acoustic scale, $\ell_A$. 
We then showed how one of the largest parameter degeneracy in the CMB data
could be understood in terms trading off $h$ and $\Omega_m$ in $\theta_A$.
Finally, we showed that \WMAP's measure of the age of the universe
is largely a function of \WMAP's identification of
the position of the first peak. 

\subsection{The Amplitude of the First Peak}

The scaling of the amplitude of the first peak with $\omega_b$ and
$\omega_m$ has a straightforward interpretation. However, unlike
the position, the amplitude itself has a complicated
dependence on the cosmological parameters 
\citep{efstathiou/bond:1999,hu/etal:2001}. We defer discussing
the $\omega_b$ scaling until after \S\ref{sec:second_peak}
and consider here just the $\omega_m$ scaling. 

Increasing $\omega_m$ decreases the first peak height.
When the universe is radiation dominated at $z>z_{eq}$ and 
the photon-baryon fluid compresses in a gravitational
potential well, the depth of the well is decreased due to the 
additional mass loading of the fluid. As a result, the compressed
fluid can expand more easily because it sees a 
shallower potential. The process is termed ``radiation driving''
\citep{hu/sugiyama:1995,hu/dodelson:2002} and enhances
the oscillations. (In addition, a shallower potential well
means that the photons are redshifted less as they climb out
enhancing the effect further.)
The effect remains important through recombination.
Increasing $\omega_m$, while holding all else but $\Omega_\Lambda$ constant, 
moves the epoch of matter-radiation 
equality to higher redshifts without significantly affecting $z_{dec}$. 
This gives the dark matter more time
to develop the potential wells into which the photon-baryon fluid 
falls. With better defined potential wells, the
compressed photon-baryon fluid has less of an effect on them,
thereby reducing the effects of radiation driving.
As a consequence, increasing $\omega_m$ decreases the 
amplitude of the fluctuations for all the acoustic peaks. 
 
\subsection{Additional TT Peaks and Troughs\label{sec:second_peak}}

There is a pronounced second peak in the {\WMAP} TT spectrum
at a height $>5\sigma$ above the first trough.
The peak arises from the rarefaction phase of an acoustic
wave. In broad terms \citep{hu/dodelson:2002}, in the same conformal time that it takes
the plasma to compress over the acoustic horizon, an acoustic wave
with half the wavelength (twice the $\ell$) of the first peak
can compress and rarify. Likewise, a compressional third peak
is expected at the second harmonic of the first peak and a
rarefaction fourth peak is expected at the third harmonic of the 
first peak. This set of peaks, the first two of which {\WMAP}
clearly sees, shows that the photon-baryon fluid underwent
acoustic oscillations.  

The position of the first trough, and the second and subsequent
peaks is predicted from the acoustic scale and phase shifts.
From equation~(\ref{eq:peak_phase}) and the values in Table~\ref{tab:numvals},
one predicts
$\ell_{1.5}^{TT}=\dlpredictedlthreehalfs$,
$\ell_2^{TT}=\dlpredictedltwo$,
and $\ell_3^{TT}=\dlpredictedlthree$.
The first two of these values agree
with those found from the peak fits to within the measurement 
uncertainty as shown in Table~\ref{tab:peaks}.

The amplitude of the second peak depends on many of the same parameters
as the amplitude of the first. Its most distinctive feature, though,
is that increasing $\omega_b$ {\it decreases} its height.
The reason is that as one increases the baryon density one increases
the inertia in the photon-baryon fluid. When compared to the lower
$\omega_b$ case, the compressions are deeper and the rarefactions 
not as pronounced. Thus, the compressional peaks are hotter (first
and third) and the rarefaction peaks (second and fourth) are cooler. 

Following \cite{hu/etal:2001} we characterize the amplitude of
second peak as a ratio to the amplitude of the first peak, $H_2^{TT}$.  
The ratio is insensitive to the reionization optical depth and to the overall
amplitude of the spectrum since they simply scale the amplitudes of both peaks. 
It depends on just $\omega_b$, $\omega_m$, and $n_s$. The dependence
on $\omega_m$ is relatively weak, since to a first approximation it
too just scales the peak amplitudes as discussed above.
The dependence on $n_s$ simply comes from the overall slope of the CMB
angular spectrum. Increasing $n_s$ increases the height of the second
peak relative to the first. The following function is derived from
fitting to a grid of CMBFAST spectra. It gives the ratio of the 
peak amplitudes to $2$\% accuracy for 50\% variations in the parameters:
\begin{eqnarray}
H^{TT}_2 \equiv 
\left({{\Delta T^{\rm TT}_{2}}\over{\Delta T^{\rm TT}_{1}}} \right)^2
        &= & 0.0264\, \omega_b^{-0.762} (2.42)^{n_s-1} \nonumber \\
        &&\times e^{-0.476 \ln(25.5\omega_b + 1.84\omega_m)^2}
\label{eq:h2tt}
\end{eqnarray}
the parameter dependences are 
\begin{eqnarray}
{\Delta  H^{TT}_2 \over  H^{TT}_2  }=
0.88\Delta n_s
- 0.67{\Delta \omega_b \over \omega_b }
+ 0.039{\Delta \omega_m \over  \omega_m }.
\end{eqnarray}
Thus a 1\% increase in both $\omega_b$ and $\omega_m$
with $n_s$ fixed
reduces the height of the second peak relative to the first
by 0.63\%. 

\begin{figure*}[tb]
\plottwo{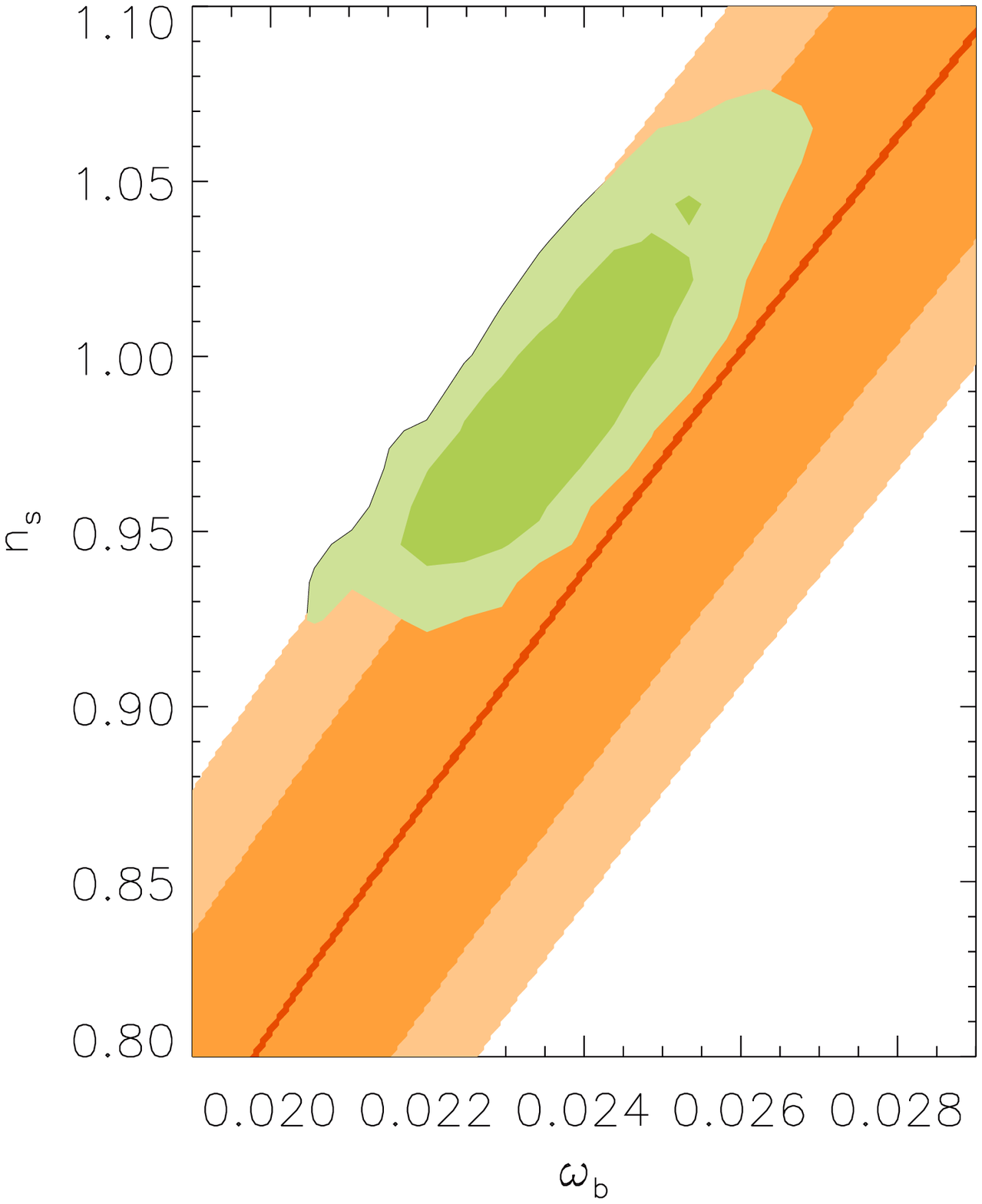}{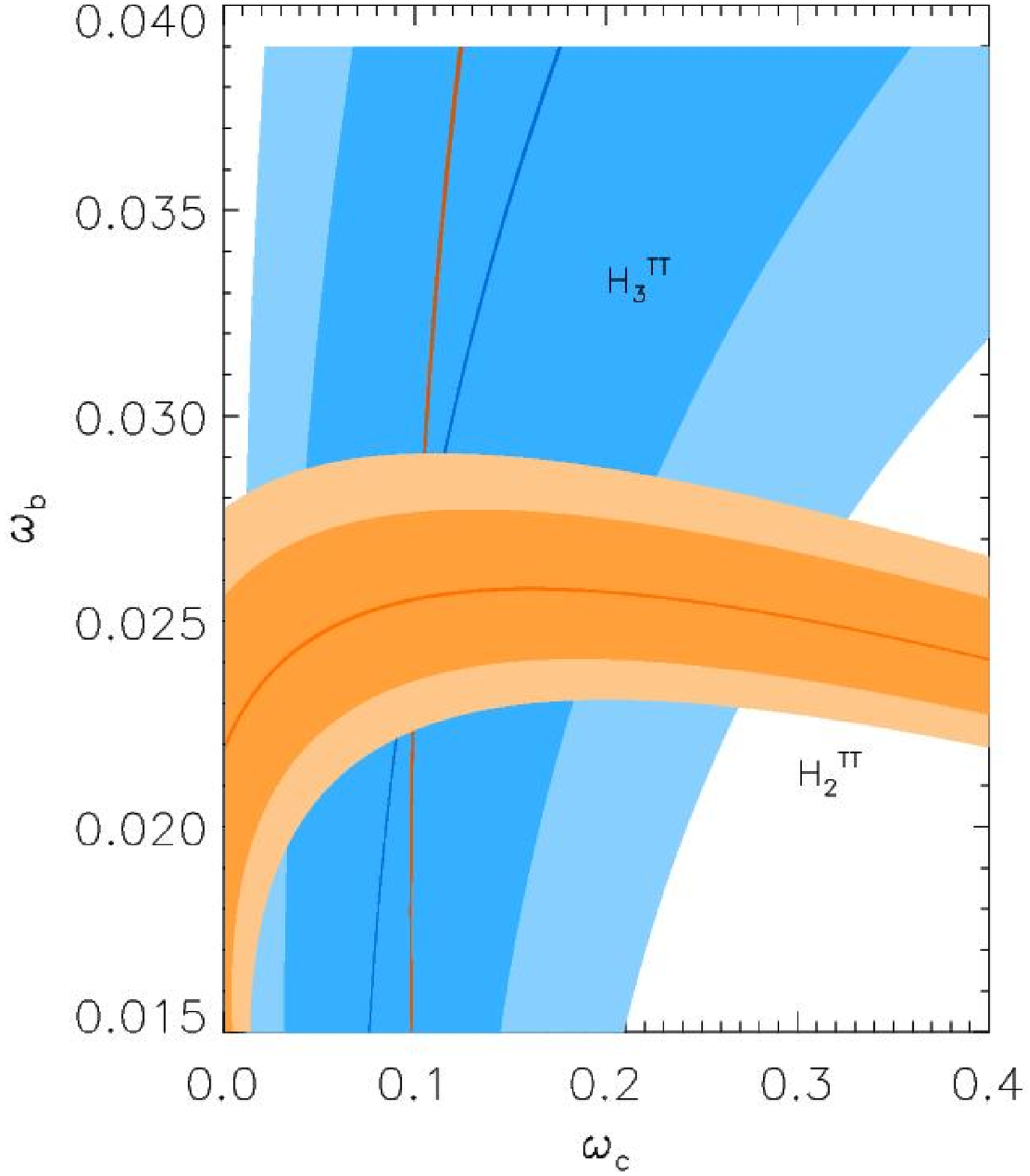}
\caption{Left: Parameter restrictions from $H_2^{TT}$ in the 
$\omega_b-n_s$ plane. The orange swath is the $1\sigma$ band 
corresponding to $H_2^{TT}=\tthtwo$ with
$\omega_m=\ensuremath{0.14}$. The swath is broadened
if one includes the uncertainty in $\omega_m$.
The light orange swath is $2\sigma$. The solid line in the middle of the 
swath is for $\Delta H_2^{TT}= \Delta \omega_m =0$. The green contours are
from the full analysis of just the {\WMAP} data
and are thus more restrictive. Right:
The constraints in the $\omega_b-\omega_c$ plane from the peak 
ratios in a flat geometry with $n_s=\ensuremath{0.99}$. 
The dark shaded regions in each swath are the $1\sigma$
allowed range; the light shaded regions show the $2\sigma$ range.
Orange is for $H_2^{TT} = \tthtwo$,
blue is for $H_{3}^{TT} = \tththree$, and red is for
$H_{2}^{TE} = \tehonehalf$. The uncertainty band for $H_{2}^{TE}$
is not shown as it is broader than the $H_{3}^{TT}$ swath.
The heavier central lines correspond to $\Delta H_2^{TT}=0$, 
$\Delta H_3^{TT}=0$, and $\Delta H_2^{TE}=0$, each with $\Delta n_s=0$. 
As the mission progresses, all uncertainties will shrink.}
\label{fig:n_wb_wc}
\end{figure*}

For the {\WMAP} data, $H^{TT}_2=\tthtwo$.
In Figure~\ref{fig:n_wb_wc} we show the constraints from 
$H^{TT}_2$ for $\omega_m = \ensuremath{0.14}$.
The contour lines correspond to the full analysis \citep{spergel/etal:2003}. 
One sees that the two analyses
are consistent, though the error from just considering $H^{TT}_2$
is larger. 
However, it is clear that the uncertainty in $H_2^{TT}$ for
$\omega_b$ at fixed $n_s$ and $\omega_m$ leads to 
nearly the same uncertainty as deduced from the full analysis.
Thus, we may interpret \WMAP's ability to determine
$\omega_b$ as coming primarily from the precise measurement of the ratio
of the first two peaks. The first two terms on the right side of the 
above equation, as shown in Figure~\ref{fig:n_wb_wc},  
also quantify the $\omega_b-n_s$ degeneracy \citep{spergel/etal:2003}.

The height of the third peak {\it increases} as $\omega_b$ increases,
as discussed above, and increasing $\omega_m$ decreases its height.
Thus the ratio to the first peak is not as distinctive as for the 
second peak in terms of $\omega_b$ and $\omega_m$ but the
long $\ell$ baseline makes the ratio more sensitive to $n_s$. 
\cite{hu/etal:2001} give: 
\begin{eqnarray}
H_3^{TT} &\equiv& \left({ \Delta T^{\rm TT}_{3} 
                \over \Delta T^{\rm TT}_{1} }\right)^2 \\ \nonumber
&=& {2.17\omega_m^{0.59}\, 3.6^{n_s-1}\over 
      [1+(\omega_b/0.044)^2][1+1.63(1-\omega_b/0.071)\omega_m]}
\label{eq:h3tt}
\end{eqnarray}
with parameter dependencies
\begin{eqnarray}
{\Delta  H^{TT}_3 \over  H^{TT}_3  }=
1.28\Delta  n_s
-0.39{\Delta  \omega_b \over \omega_b  }+
0.46{\Delta  \omega_m \over  \omega_m  }.
\end{eqnarray}
With $n_s$ fixed, increasing $\omega_b$ and $\omega_m$ by 1\%
increases the height of the third peak relative to the first
by only 0.07\%. Measuring the third peak helps mostly
in measuring $n_s$, thereby breaking the $\omega_b-n_s$ degeneracy
shown in the left side of Figure~\ref{fig:n_wb_wc}.
For {\WMAP} parameters, equation~\ref{eq:h3tt} is
accurate to 1\%.

{\WMAP} does not yet clearly measure the third peak.
For its height, we use the value of the \cite{wang/etal:2002} compilation
because it is the most recent and includes calibration error.
At $l=801$,  $(\Delta T_\ell^{\rm TT})^2=2322\pm440\,\uK^2$. Though not 
quite at the peak, the value is sufficient. With \WMAP's first 
peak, $H_3^{TT} = \tththree$.  

\subsection{The Temperature-Polarization peaks\label{sec:p_amp}}

The E-mode polarization in the CMB arises from Thompson scattering of 
a quadrupolar radiation pattern at the surface of last scattering.
At decoupling, the quadrupolar pattern is produced by 
velocity gradients in the plasma 
and is correlated with the temperature anisotropy 
\citep{bond/efstathiou:1984,coulson/crittenden/turok:1994}. 
In rough terms, because it is velocity dependent,
the temperature-polarization correlation traces the derivative 
of the temperature spectrum.
For a given TT spectrum, the (non-reionized) TE and EE spectra are predicted.
Thus, their observation confirms that the cosmological model is correct 
but in so far as they are predicted for $\ell>40$,
they contain little additional information about the parameters. 
By contrast, the TE polarization signal at $\ell<20$ \citep{kogut/etal:2003},
produced by reionized electrons scattering the local
$z_r\sim\zreion$ CMB quadrupole, breaks
a number of parameter degeneracies and considerably enhances
the ability to extract the cosmic parameters.

From parameterizing the output of CMBFAST, we find that the 
ratio of the first TE antipeak to the second TE peak is given by
\begin{eqnarray}
\lefteqn{H^{\rm TE}_{2} \equiv
         -\left({{\Delta T^{\rm TE}_{1}}
                \over{\Delta T^{\rm TE}_{2}}}\right)^2
        = 0.706\, \omega_m^{0.349} (0.518)^{n_s-1}}
        \hspace{2.5truein}\nonumber\\
        \, \times e^{0.195 \ln(33.6\omega_b + 5.94\omega_m)^2}.
\end{eqnarray}
The parameter dependencies are
\begin{eqnarray}
{\Delta  H^{TE}_{2} \over  H^{TE}_{2}  }=
-0.66\Delta  n_s +
0.095{\Delta  \omega_b \over \omega_b  }+
0.45{\Delta  \omega_m \over  \omega_m  }.
\end{eqnarray}
The sign preceding $\Delta n_s$ is opposite to that in the case
of the TT peaks since we are considering the ratio
of the first antipeak (low $\ell$) to the second peak (high $\ell$),
whereas in the case of the TT peaks we were considering the
second peak (high $\ell$) to the first peak (low $\ell$).
Here, increasing $\omega_m$ {\it increases} the contrast between the 
TE peak and antipeak. This occurs because 
increasing $\omega_m$ leads to deeper potential wells at decoupling.
In turn, this produces higher velocities and thus an enhanced 
polarization signal. From the data we find $H^{\rm TE}_{2}=\tehonehalf$. 
 
\subsection{Combined Peak Constraints}

We now use the scaling relations just presented in a 
flat model with the scalar index limited to
$n_s=\ensuremath{0.99}$. 
This simplified minimal model leads to a greater
intuition for assessing the peaks' role in determining 
the cosmological parameters. 
The constraints from $H_2^{TT}$ and $H_3^{TT}$
are shown in Figure~\ref{fig:n_wb_wc}.
We see that $H_2^{TT}$ determines $\omega_b$
for most values of the dark matter and that $\omega_b<0.029$
(2$\sigma$) regardless of the amount of dark matter.
The width of the swath is about twice that of the formal error,
again indicating that more in the spectrum than just $H_2^{TT}$ constrains
$\omega_b$. 

The constraints from $H_{2}^{TE}$ and $H_3^{TT}$ are 
almost orthogonal to the $H_2^{TT}$ constraint but their 
uncertainty bands are wide. This shows us that
the {\WMAP} value for $\omega_c$ (or $\omega_m$) is not being driven 
specifically by the peak ratios. Rather, the constraint comes
from the amplitude of the whole spectrum. The degeneracy with 
$\tau$ is broken by the TE detection of reionization and the degeneracy
with the overall amplitude is broken because $\omega_m$ 
affects primarily the acoustic peak region more than the 
low $\ell$ part of the spectrum.  

\section{Discussion\label{sec:conclude}}
 
We have focused on the dominant acoustic features in the {\WMAP} 
spectrum, but there are other features as well. 
Notable by its absence is a trough 
in the TT spectrum at $\ell\approx10$ due to the increase in fluctuation
power near $\ell=2$ from the 
integrated Sachs-Wolfe effect in a $\Lambda$-dominated universe.
This feature is largely obscured by cosmic variance. For {\WMAP} though,
there is very little power at low $\ell$ and the correlations
with $\theta>60^\circ$ are unusually small \citep{bennett/etal:2003b}.
The lack of even a hint of an upturn at low $\ell$,
in the context of $\Lambda$CDM models, is surprising. 
(However, {\COBE} also found no evidence for such an upturn). 
This departure from the model constitutes a small (though possibly
important) fraction of the total fluctuation power and does
not cast doubt on the interpretation of the $\ell>40$ spectrum.

At $\ell\approx 40$ and $\ell\approx 210$, for example,  
there are excursions from a smooth spectrum that are somewhat 
larger than expected statistically. While intriguing,
they may result from a combination of cosmic variance, subdominant
astrophysical processes, and small effects from approximations
made for the year one data analysis \citep{hinshaw/etal:2003}.
At present, we consider them interesting
but do not attach cosmological significance to them
\citep[e.g.,][]{peiris/etal:2003,bose/grishchuk:2002}.
More integration time and more detailed analyses are needed
to understand how they should be interpreted.

In summary, the characteristics of the peaks of the {\WMAP}
angular power spectra may be robustly extracted from the data.
The second TT peak is seen with high accuracy; the TE antipeak and peak
have been observed for the first time. In the context of a flat
adiabatic $\Lambda$CDM model, which fits the {\WMAP} data well,
we report the characteristics of the decoupling surface.
By considering a reduced parameter space, we show how the position 
of the first peak leads to \WMAP's tight determination of the age
of the universe and how $\omega_b$ is determined primarily 
from the amplitude ratio of the first to second TT peak.
The determination of $\omega_m$ comes from considering
the full data set and is not attributable to any particular
peak ratio.

\acknowledgments

LP is grateful to Viatcheslav Mukhanov for helpful discussions on the 
parameter dependences of the peaks during his year at Princeton 
and to Alessandro Melchiorri for discussing his work on peaks 
prior to publication. LP especially thanks 
Toby Marriage for many enlightening discussions on the first peak.
We also thank Eiichiro Komatsu, Licia Verde, Will Sutherland,
and an anonymous reviewer for comments that improved this paper.
The {\WMAP} mission is made possible by the support of the Office of Space 
Sciences at NASA Headquarters and by the hard and capable work of scores of 
scientists, engineers, technicians, machinists, data analysts, 
budget analysts, managers, administrative staff, and reviewers.

\begin{deluxetable}{cccccc}
\tabletypesize{\small}
\tablecaption{Previous Measurements of the First Peak\label{tab:others_peaks}}
\tablehead{
\colhead{Experiment} &
\colhead{$\ell_1^{TT}$} &
\colhead{$\Delta T_{1,TT}$ [$\mu$K]} &
\colhead{Cal. error} &
\colhead{$\ell$ range} &
\colhead{Reference}
}
\startdata
TOCO  & $207^{+15}_{-12}$ & $87^{+9}_{-8}$ &  8\%  & $60<\ell<410$ 
& \cite{miller/etal:1999}   \\
BOOMERANG-NA & $207^{+26}_{-20}$ & $69.6^{+9.7}_{-8.4}$ & 8\% & $50<\ell<410$ 
& \cite{mauskopf/etal:2000} \\
MAXIMA  & $236\pm14$  & $70.0^{+5.3}_{-5.3}$ & 4\% & $75<\ell<385$ 
& \cite{lee/etal:2001}    \\
DASI  & $199^{+15}_{-18}$ & $71.9\pm4.4$ & 4\% & $110<\ell<380$   
& \cite{halverson/etal:2002}  \\
VSA  & $236^{+20}_{-26}$  & $73.9\pm6.8$ & 3.5\% & $140<\ell<420$ 
& \cite{grainge/etal:2003} \\
BOOMERANG &$219\pm5$  & $73.8^{+8.5}_{-7.0}$ & 10\% & $100<\ell<410$ 
& \cite{ruhl/etal:2003}\\
ARCHEOPS & $220^{+7}_{-6}$ & $70.2^{+5.7}_{-4.9}$ & 7\% & $18<\ell<330$ 
& \cite{benoit/etal:2003}  \\[1pt]
\hline
{\WMAP}  & \ttponel & \ttponeh & 0.5\% & $100<\ell<350$ & This paper  \\
\enddata
\tablecomments{
All values come from fitting a Gaussian shape to just the first peak
of the data set specified, and include calibration error.
Each data set is considered on its own, without the \COBE/DMR
data, and so a direct comparison between experiments may be made. 
The best fit position depends somewhat on the fitting function
so the values from different analyses yield different results
\citep[e.g.,][]{knox/page:2000,durrer/novosyadlyj/apunevych:2003,
odman/etal:2002,grainge/etal:2003}. 
The TOCO, VSA, and BOOMERANG-NA experiments were calibrated with Jupiter.
The TOCO and VSA experiments are most affected because of they operate 
at 30-150 GHz and 35 GHz respectively. With the new calibration of 
Jupiter \citep{page/etal:2003}, the peak values above will be reduced
$\approx 5\%$.
The weighted peak amplitude is $71.7\pm2.4~\mu$K and the 
weighted peak position is $218.8\pm3.5$
in good agreement with {\WMAP}. In a separate analysis based on 
different assumptions, Bond
reports $\ell_1^{TT}=222\pm3$ (private communication). This
was also the value preferred by a concordance model \citep{wang/etal:2000}
that predated all the experiments of the new millennium.
Note that {\WMAP}'s values for the position and amplitude
are both more than four times more precise than all the listed 
measurements combined.} 
\end{deluxetable}

\begin{deluxetable}{ccccccc}
\tabletypesize{\small}
\tablecaption{{\WMAP} Peak and Trough Amplitudes and Positions\label{tab:peaks}}
\tablehead{
\colhead{Quantity} &
\colhead{Symbol} &
\colhead{$\ell$} & 
\colhead{$\Delta T_\ell$ [$\uK$]} &
\colhead{$\Delta T^2_\ell$ [$\uK^2$]} & 
\colhead{FULL $\ell$} &
\colhead{FULL $\Delta T^2_\ell$ [$\uK^2$]}
}
\startdata
First TT Peak & $\ell_1^{TT}$ & \ttponel & \ttponeh & \ttponehsq & \mcmcttponel & \mcmcttponehsq \\
First TT Trough & $\ell_{1.5}^{TT}$ & \tttonel & \tttoneh & \tttonehsq & \mcmctttonel & \mcmctttonehsq \\
Second TT Peak & $\ell_2^{TT}$ & \ttptwol & \ttptwoh & \ttptwohsq & \mcmcttptwol & \mcmcttptwohsq \\
First TE Antipeak & $\ell_{1}^{TE}$ & \tetonel & \nodata & \tetonehsq & \mcmctetonel & \mcmctetonehsq \\
Second TE Peak & $\ell_2^{TE}$ & \teptwol & \nodata & \teptwohsq & \mcmcteptwol & \mcmcteptwohsq \\
\enddata
\tablecomments{
The values and uncertainties are the maximum and 
width of the a~posteriori distribution of
the likelihood assuming a uniform prior. 
The uncertainties include calibration uncertainty and cosmic
variance.
The FULL values are derived from the full CMBFAST-based likelihood analysis
using just the {\WMAP} data \citep{spergel/etal:2003}.
The FULL method yields consistent results. Recall that
the FULL chains are sensitive to the combined TT and TE spectra and not 
just the individual peak regions.
Numerical errors in CMBFAST will increase the uncertainties, but should
not bias the results.} 
\end{deluxetable}

\begin{deluxetable}{lccc}
\tabletypesize{\small}
\tablecaption{{\WMAP} Cosmological Parameters for the Peaks Analysis\label{tab:numvals}}
\tablehead{
\colhead{Quantity} & \colhead{Symbol} & \colhead{Value} & 
}
\startdata
Physical baryon density & $\omega_b$ & \ensuremath{0.024 \pm 0.001} & \\
Physical mass density & $\omega_m$ & \ensuremath{0.14 \pm 0.02}  & INPUT \\
Scalar index & $n_s$ & \ensuremath{0.99 \pm 0.04} &\\ 
\noalign{\smallskip\hrule\smallskip}
First TT peak phase shift & $\phi_1$ & \dlphione & \\
First TT trough phase shift & $\phi_{1.5}$ & \dlphithreehalfs & \\
Second TT peak phase shift & $\phi_2$ & \dlphitwo & DERIVED  \\
Third TT peak phase shift & $\phi_3$ & \dlphithree &FROM \\
Redshift at decoupling & $z_{dec}$ & \ensuremath{1088^{+ 1}_{- 2}} & INPUT \\
Redshift at matter radiation equality & $z_{eq}$ & \ensuremath{3213^{+ 339}_{- 328}} &\\ 
Comoving acoustic horizon size at decoupling (Mpc) & $r_s$ & \rsdec &\\
\noalign{\smallskip\hrule\smallskip}
Acoustic scale & $l_A$ & \acousticscale & DERIVED FROM  \\ 
Comoving angular size distance to decoupling (Gpc) & $d_A$ 
& \dAdec & INPUT + PEAKS \\ 
\enddata
\tablecomments{
The cosmological parameters in the top section are derived
from just the {\WMAP} data 
assuming a flat $\Lambda$CDM model \citep{spergel/etal:2003}.
The quantities in the middle section are derived from the cosmological
parameters in the top section. The quantities in bottom section are calculated
using the middle quantities and the measured position of the first
peak.
The quantity
$z_{dec}$ which we use corresponds to the location of the maximum 
of the visibility function in CMBFAST. The quantity
computed using \cite{hu/sugiyama:1996} corresponds to
$\tau(z_{dec})=1$ and is $1090\pm2$.
}
\end{deluxetable}

\end{document}